\documentclass[journal]{IEEEtran}

\usepackage{graphicx}
\usepackage{epstopdf}
\usepackage{amsmath}
\usepackage{amsfonts,amssymb}
\usepackage{float}
\usepackage{subfigure}
\usepackage{algorithmic}
\usepackage{algorithm}
\usepackage{booktabs}
\usepackage{soul, color, xcolor}
\newcommand{\mathcolorbox}[2]{\colorbox{#1}{$\displaystyle #2$}}
\ifCLASSINFOpdf
\else
\usepackage[dvips]{graphicx}
\fi
\usepackage{url}

\begin{document}
	
	\title{Three-Dimensional Sparse Random Mode Decomposition for Mode Disentangling with Crossover Instantaneous Frequencies}
	
	\author{Chen Luo, Tao Chen, and Lei Xie, and Hongye Su, \IEEEmembership{Senior Member, IEEE}
		\thanks{This work was supported in part by the National Key R\&D Program of China (No. 2022YFB3305900) and the National Natural Science Foundation of P.R. China (NSFC: 62073286) (\textit{Chen Luo and Tao Chen contributed
				equally to this work.}) (\textit{Corresponding author: Tao Chen; Lei Xie.})}
		\thanks{Chen Luo, Tao Chen, Lei Xie and Hongye Su are with the State Key Laboratory of Industrial Control Technology, Zhejiang University, Hangzhou 310027, China (e-mail: lawchen333@gmail.com; chentao227722@163.com; leix@iipc.zju.edu.cn; hysu@iipc.zju.edu.cn}
	}
	
	\markboth{Journal of \LaTeX\ Class Files, Vol. 14, No. 8, August 2015}
	{Shell \MakeLowercase{\textit{et al.}}: Bare Demo of IEEEtran.cls for IEEE Journals}
	\maketitle
	
	\begin{abstract}
		Sparse random mode decomposition (SRMD) is a novel algorithm that constructs a random time-frequency feature space to sparsely approximate spectrograms, effectively separating modes. However, it fails to distinguish adjacent or overlapped frequency components, especially, those with crossover instantaneous frequencies. To address this limitation, an enhanced version, termed three-dimensional SRMD (3D-SRMD), is proposed in this letter. In 3D-SRMD, the random features are lifted from a two-dimensional space to a three-dimensional (3D) space by introducing one extra chirp rate axis.
		This enhancement effectively disentangles the frequency components overlapped in the low dimension.
		Additionally, a novel random feature generation strategy is designed to improve the separation accuracy of 3D-SRMD by combining the 3D ridge detection method. Finally, numerical experiments on both simulated and real-world signals demonstrate the effectiveness of our method.
	\end{abstract}
	

	\IEEEpeerreviewmaketitle

	\section{Introduction}
	
	\IEEEPARstart{N}{on-stationary} signals are ubiquitous in both natural  \cite{hadjidimitriou2012toward,chen2023sinusoidal} and engineering systems  \cite{feng2013recent,wang2018matching}. These signals are typically modeled as superpositions of amplitude and frequency-modulated modes, called multi-component signals (MCSs)  \cite{auger2013time}. To reveal the time-varying characteristics of these signals, time-frequency analysis (TFA) methods are introduced. A significant challenge within the TFA area is to separate the intrinsic modes within MCSs, commonly referred to as signal decomposition.
	
	Initially, most decomposition methods directly extract intrinsic mode from MCSs in the time domain, e.g., the empirical mode decomposition \cite{huang1998empirical} and its variants \cite{ wu2009ensemble, yeh2010complementary}. However, these time-domain methods always lack mathematical foundations.
	On the other hand, since many real-world signals exhibit sparsity in the Fourier spectrum, variational mode decomposition \cite{dragomiretskiy2013variational} and its improved versions \cite{nazari2020successive, nazari2017variational} have been proposed to separate the intrinsic modes from MCSs in the frequency domain.
	However, these methods formulated in the frequency domain are unable to extract the wide-band modes that have an overlapping spectrum.
	Recently, to analyze wide-band signals, Chen et al. developed two advanced methods based on the multi-component chirp signal model, i.e., nonlinear chirp mode decomposition (NCMD) \cite{chen2017nonlinear} and intrinsic chirp component decomposition (ICCD) \cite{chen2017intrinsic}.
	
	More recently, sparse random mode decomposition (SRMD) was proposed as a novel signal decomposition method \cite{richardson2023srmd}. Inspired by the sparse random feature expansion \cite{hashemi2023generalization,saha2023harfe}, SRMD begins by assuming that a signal can be approximately represented as the sum of sparse random time–frequency (TF) features. 
	A spatial clustering algorithm is then utilized to separate the localized random features, thereby effectively achieving mode separation with less mode mixing and fewer Gibbs phenomena. However, adjacent or overlapped frequency components cannot be separated in the two-dimensional (2D) random feature space. Thus, SRMD is unable to disentangle modes with crossover instantaneous frequencies (IFs). 
	
	\hl{To address this issue, an improved method called three-dimensional SRMD (3D-SRMD) is proposed.
	Motivated by the chirplet transform (CT)} \cite{mann1995chirplet,li2022chirplet,chui2023analysis}, \hl{we lift the random features from the TF plane to a three-dimensional (3D) space, i.e., time-frequency-chirprate (TFC).} This enhancement effectively disentangles the frequency components overlapped in the low dimension.
	\hl{Additionally, unlike SRMD, the concentrated generation of random features ensures the sparsity of the 3D random feature space and eliminates the dependence on clustering algorithms.}
	By combining the 3D ridge detection (RD) method, this new random feature generation strategy effectively improves the mode separation of 3D-SRMD.

	\section{Sparse Random Mode Decomposition}
	The SRMD is achieved on the sparse random feature approximation to the inverse short-time Fourier transform (STFT). Specifically, a signal $x(t) \in L^{1}(\mathbb{R})$ can be represented by: \begin{equation}\label{eq1}
		x(t)=\int_{-\infty}^{+\infty} \int_{-\infty}^{+\infty} F_{x}^{g}(\tau, \xi) g(t-\tau) e^{j 2 \pi \xi t} d \tau d\xi,	
	\end{equation}
	where $F_{x}^{g}(\tau, \xi)$ denotes the STFT of $x(t)$ with $g(t)$; $g(t)$ is a (positive) window function such that $\int_{-\infty}^{+\infty} g(\tau) d \tau=1$. 
	
	Employing the sparse random feature expansion, the signal can be approximated by random features \cite{hashemi2023generalization}:
	\begin{equation}\label{eq2}
		\begin{aligned}
			x(t)&=\int_{-\infty}^{+\infty} \int_{-\infty}^{+\infty} F_{x}^{g}(\tau, \xi) g(t-\tau) e^{j 2 \pi \xi t} d \tau d\xi, \\	
			&\approx \sum_{i=1}^{N} c_{i} g(t-\tau_{i}) e^{j 2 \pi \xi_{i} t}= \sum_{i=1}^{N} c_{i}  \varphi _i(t),
		\end{aligned}
	\end{equation}
	where $(\tau_1,\xi_1), \cdots , (\tau_N,\xi_N)$ are drawn independently and identically distributed (i.i.d.) from a distribution $p$ ($p$ is chosen to be uniform in SRMD); $\varphi_i(t)$ represents the random feature; $c_i \in \mathbb{C}$ denotes the weight coefficients of $ \varphi _i(t)$.
	
	Reference \cite{rahimi2008weighted} proved that, for any $\delta > 0$, with probability at least $1 - \delta$ over $(\tau_1,\xi_1), \cdots , (\tau_N,\xi_N)$, there exist $c_i$ such that the following approximation error bound in (\ref{eq2}) holds:
	
	\begin{equation}\label{error}
		\begin{aligned}
			\Vert x(t)-\sum_{i=1}^{N} c_{i}  \varphi _i(t) \Vert_2 \leq \frac{C}{\sqrt{N}}(1+\sqrt{2 \log \frac{1}{\delta}}),
		\end{aligned}
	\end{equation}
	where $C=\sup \frac{\mid F_{x}^{g} (\tau, \xi) \mid}{p(\tau, \xi)}$. 
	This theory ensures the existence of $c_i$, paving the way for learning of optimal $c_i$ through the construction of basis pursuit de-noising (BPDN) problem \cite{chen2001atomic,figueiredo2007gradient} and its solution via the L1 norm spectral projection gradient (SPGL1) algorithm \cite{van2009probing,van2011sparse}.
	
	In SRMD, $(\tau_i,\xi_i)$ with non-zero $c_i$ obtained by the BPDN is expected to form a sparse TF representation of $x(t)$. These pairs with non-zero coefficients are then clustered using the density-based spatial clustering of applications with noise (DBSCAN) algorithm \cite{ester1996density}, thereby enabling the reconstruction of each mode based on the grouping of clusters, as follows:
		\begin{equation}\label{rst}
		\begin{aligned}
			x_k(t) = \sum_{i \in I_k} c^*_{i}  \varphi _i(t), k=1, \cdots, K,
		\end{aligned}
	\end{equation}
	where $c^*_{i}$ is the coefficient obtained by solving the BPDN problem; $K$ is the number of groups in the clustering results; $I_k$ represents the index set of the $k$-th group of random features with non-zero coefficients.

	\begin{figure}[H]
	\centering
	\begin{minipage}[t]{0.5\linewidth}
		\centering
		\includegraphics[width=1.75in]{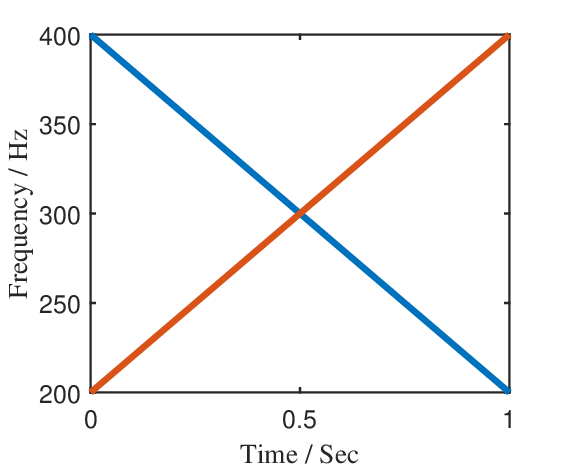}
		%\caption{fig1}
	\end{minipage}%
	\begin{minipage}[t]{0.5\linewidth}
		\centering
		\includegraphics[width=1.75in]{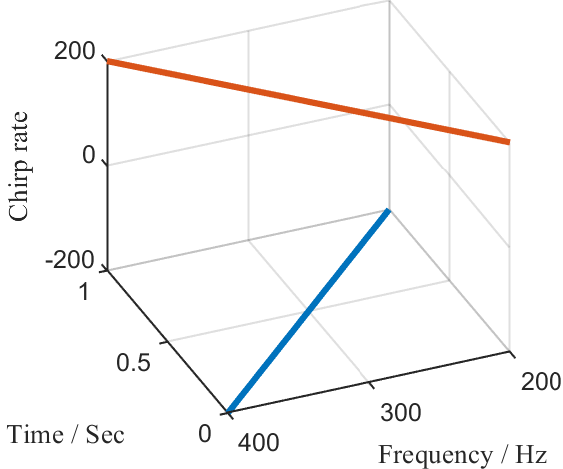}
		%\caption{fig2}
	\end{minipage}%
	
	\centering
	\caption{The specific crossover frequency components $s_1=\cos(2\pi(400t-100t^2))$ and $s_2=\cos(2\pi(200t+100t^2))$ in (\textbf{left}) TF plane, (\textbf{right}) TFC space.}
	\label{3dplane}
\end{figure}	
	
\label{key}	\section{Proposed Method}
	\label{sec:guidelines}
	
	\subsection{Three-Dimensional Random Feature Space}
	In (\ref{eq2}), a specific pair of TF parameters $(\tau_i,\xi_i)$ uniquely determines a feature $ \varphi _i(t)$.
	When the signal contains overlapped frequency components, the random features of different modes in the overlapped region have the same TF parameters. Consequently, it is impossible to separate the modes with crossover IFs in the 2D random TF feature space.
	Inspired by CT \cite{mann1995chirplet,li2022chirplet,chui2023analysis}, we introduce chirp rate (CR) parameter to lift the random feature to TFC space. \hl{CR represents the rate at which frequency changes over time. When modes exhibit frequency crossover, they typically have distinct CRs at the crossover moment.} Therefore, signals with overlapping frequency components, which entangle in 2D TF plane, can be separated in 3D TFC space (see Fig. \ref{3dplane} for an illustration). 
	
	Incorporating the CR into $ \varphi _i(t)$ of (\ref{eq2}), we obtain the 3D random features, as follows:
	\begin{equation}\label{eq5}
		\hat{\varphi}_i(t) = g(t-\tau_{i}) e^{j 2 \pi \xi_{i} t} e^{j \pi \beta_{i} (t-\tau_{i})^2},
	\end{equation}
	where $\beta_i$ represents the newly introduced random chirp rate parameter.
	Similar to $ \varphi _i(t)$, the random feature $ \hat{\varphi} _i(t)$ corresponds to a specific tuple of TFC parameters $(\tau_i,\xi_i, \beta_{i})$.
	Subsequently, a multitude of $\hat{\varphi}_i(t)$ constructs a 3D random feature space. 
	Due to differences in TFC parameters, overlapped frequency components can be distinguished within this 3D space, ensuring the feasibility of separating modes with crossover IFs.

	\subsection{Concentrated Distribution of Random Features}
\begin{figure}[H]
		\centering
		\begin{minipage}[t]{0.5\linewidth}
			\centering
			\includegraphics[width=1.75in]{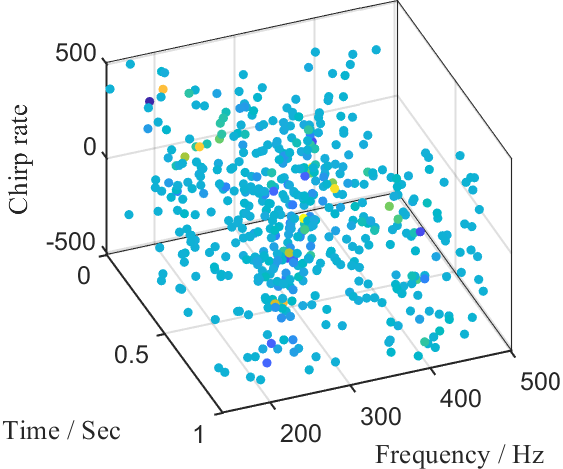}
			%\caption{fig1}
		\end{minipage}%
		\begin{minipage}[t]{0.5\linewidth}
			\centering
			\includegraphics[width=1.75in]{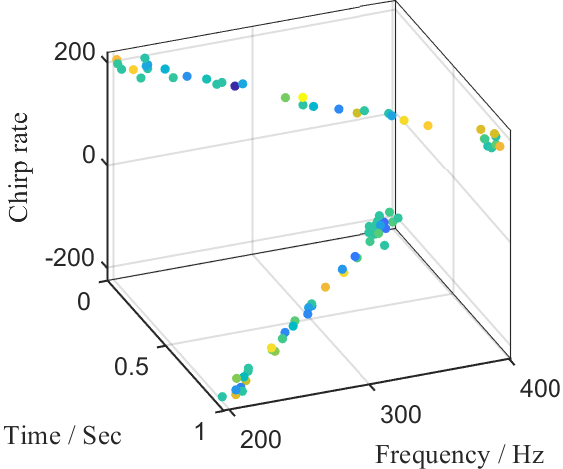}
			%\caption{fig2}
		\end{minipage}%
		
		\centering
		\caption{The 3D random feature space of the signal in Fig. \ref{3dplane} with (\textbf{left}) uniform distribution and (\textbf{right}) concentrated distribution (color intensity indicates weight magnitude).}
		\label{dis}
	\end{figure}

	Despite the enhancement in dimension should work theoretically, mode separation still fails due to the poor sparsity in the 3D random feature space (see Fig. \ref{dis} (\textbf{left})) for an illustration). Therefore, we propose a concentrated distribution strategy to solve this issue.

	Firstly, we consider a non-stationary MCS $x(t)$, defined as a superposition of AM-FM modes, as follows \cite{chen2017nonlinear}:
	\begin{equation}\label{MCS}
		\begin{aligned}
			x(t)&=\sum_{k=1}^{K} x_{k}(t)+e(t)\\
			&=\sum_{k=1}^{K} a_{k}(t) \cos (2 \pi \int_{0}^{t} f_{k}(\tau) d\tau+\phi_{k})+e(t),
		\end{aligned}
	\end{equation}
	where $t \in [0, L]$, $L$ denotes the temporal duration of the signal; $K \in \mathbb{N} $ represents the number of intrinsic modes; $a_k(t)>0$ and $f_k(t)>0$ denote the instantaneous amplitude (IA) and IF of $k$-th mode; $\phi_k$ stands for the initial phase; $e(t)$ denotes the
	additive noise.
	
	According to (\ref{eq2}), the $k$-th mode in (\ref{MCS}) can be approximated by the 2D random feature model as:
	\begin{equation}\label{mode}
		x_k(t)\approx \sum_{i=1}^{N} c_{i}^{k} g(t-\tau_{i}^{k}) e^{j 2 \pi \xi_{i}^{k} t},
	\end{equation}
	where the TF parameter pairs $(\tau_1^k,\xi_1^k), \cdots , (\tau_N^k,\xi_N^k)$ are drawn i.i.d. from a distribution $p_k$; $\left\{c_{1}^k,\cdots,c_{N}^k\right\}$ denote the weight coefficient of model, which should be identified.
	
	From (\ref{error}), the approximation accuracy of the model is influenced by the error constant $C$, which decreases as $p_k$ approximates the distribution of ${\mid F_{x_k}^{g} (\tau, \xi) \mid}$. Given that the STFT of $x_k(t)$ presents concentrated energy around its IF \cite{daubechies2011synchrosqueezed}, $x_k(t)$ can be approximated with high accuracy by the model, whose random TF features are distributed around the IFs. 
	
	Similarly, assuming that $x_k(t)$ can be approximated by the 3D random feature model from (\ref{eq5}) as:
	\begin{equation}\label{mode2}
		x_k(t)\approx \sum_{i=1}^{N} c_{i}^{k} g(t-\tau_{i}^{k}) e^{j 2 \pi \xi_{i}^{k} t} e^{j \pi \beta_{i}^{k} (t-\tau_{i})^2},
	\end{equation}
	where $(\tau_1^k,\xi_1^k,\beta_{1}^{k}), \cdots , (\tau_N^k,\xi_N^k,\beta_{N}^{k})$ are drawn i.i.d. from a distribution $q_k$. Like the 2D case, to achieve low approximation error, $q_k$ should be concentrated around the IF and CR of $k$-th mode. In this letter, $q_k$ is be considered as a band-limited uniform distribution, as follows:
	\begin{equation}\label{qk}
		\mathcolorbox{yellow}{q_k(\tau,\xi,\beta)=\begin{cases}
			\frac{1}{\lambda^2 L}, & \text{if }  \tau \in [0,L], \\
			&\xi \in [\widehat{f}_k(\tau)-\frac{\lambda}{2},\widehat{f}_k(\tau)+\frac{\lambda}{2}], \\
			&\beta \in [\widehat{f}_k^{\prime}(\tau)-\frac{\lambda}{2},\widehat{f}_k^{\prime}(\tau)+\frac{\lambda}{2}],\\
			0, & \text{else},
		\end{cases}}
	\end{equation}
	where $\widehat{f}_k(\tau)$ and $\widehat{f}_k^{\prime}(\tau)$ represent the estimated IF and CR, $\lambda$ represents the bandwidth parameter of $q_k$. \hl{Note that in this letter, 3D RD algorithm in }\cite{chen2024multiple} \hl{is applied to estimate IF and CR with the TFC representation generated by CT.} 

	Random features for modes $x_1(t),\cdots,x_k(t)$ is generated by following $q_1,\cdots,q_K$, respectively, thereby ensuring the sparsity of the 3D random feature space (see Fig. \ref{dis} (\textbf{right}) for an illustration)). Furthermore, the random features corresponding to different modes are already separated by solving BPDN, thereby obviating the need for a clustering algorithm. Similar to (\ref{rst}), the mode can be reconstructed as follows:
	\begin{equation}\label{rst2}
		\begin{aligned}
			x_k(t) = \sum \limits_{i=1}^{N} {c_{i}^k}^* \hat{\varphi}_i^k(t), k=1,\cdots,K,
		\end{aligned}
	\end{equation}
	where ${c_{i}^k}^*$ and $\hat{\varphi}_i^k(t)$ represent the coefficients and random features, which correspond to the $k$-th mode. 
	
	\subsection{Algorithm Details}
	To reduce time consumption, we provide the real field version of the random features, as follows:
	\begin{equation}\label{psi}
		\hat{\varphi}_i^k(t)= g(t-\tau_{i}^{k})\cos(2 \pi \xi_{i}^{k} t+ \pi \beta_{i}^{k} (t-\tau_{i}^{k})^2 - \frac{\pi}{2}\phi_{i}^{k}),
	\end{equation}
	\hl{where $\phi_{i}^{k} \sim \mathcal{B}(1,0.5)$ serves as a unified replacement for the cosine and sine phases, capturing the real and imaginary components of the complex random features.} Here, $g(t)=e^{-\frac{t^2}{2\alpha}}$ is chosen as a Gaussian window.
	It should be noted that, in practical applications, BPDN requires an estimation of the noise level in advance. Here, we adopt the TF segmentation algorithm of the rectified STFT in \cite{millioz2010circularity} as a noise estimator.
	
	Finally, the complete 3D-SRMD algorithm is outlined in Algorithm \ref{alg}.
	\begin{algorithm}[H]
		\caption{\hl{3D-SRMD}}
		\label{alg}
		\renewcommand{\algorithmicrequire}{\textbf{Input:}}
		\renewcommand{\algorithmicensure}{\textbf{Output:}}
		\begin{algorithmic}[1]
			\REQUIRE Signal $\boldsymbol {x} = [x_1, \cdots, x_m]^T$, sampled time $\boldsymbol {t} = [t_1, \cdots, t_m]^T$, number of random features $N$, number of modes K ,window parameters $\alpha$, bandwidth parameter $\lambda$. %%input

			\STATE \hl{Estimate the IFs $\widehat{f}_1(\tau), ... , \widehat{f}_K(\tau)$ and CRs $\widehat{f}_1^{\prime}(\tau), ...,$ $\widehat{f}_K^{\prime}(\tau)$ by 3D RD algorithm.}
			\STATE Generate distribution $q_1, ... , q_K$ as (\ref{qk})
			\FOR{$k = 1$ to $K$}
			\STATE $\left\{(\tau_i^k,\xi_i^k,\beta_i^k)\right\}_{i=1}^N$ are drawn i.i.d. from $q_k$,
			\STATE $\left\{\phi_{i}^{k}\right\}_{i=1}^N \sim \mathcal{B}(1,0.5)$.
			\ENDFOR
			\STATE Construct the random feature matrix:
			
			$\Psi=[[\hat{\varphi}_i^1(\boldsymbol{t})] \cdots [\hat{\varphi}_i^K(\boldsymbol{t})]] \in \mathbb{R}^{m \times KN}$, $\hat{\varphi}_i^k(t)$ in (\ref{psi}).
			\STATE \hl{Estimate the noise variance $\sigma^2$ of $\boldsymbol {x}$ by TF segmentation.}
			\STATE \hl{Solve the BPDN problem by SPGL1:
			
			$\boldsymbol{c}^* = \mathop{\arg\min}\limits_{\boldsymbol{c} = [c_1^1 \cdots c_N^1 \cdots c_1^K \cdots c_N^K]^T  \in \mathbb{R}^{KN}} \Vert \boldsymbol{c} \Vert_1 $
			
			$  \text{s.t.}   \Vert \Psi \boldsymbol{c} - \boldsymbol{x} \Vert_2 \leq \sqrt{m} \sigma $.}

			\ENSURE $K$ modes    %%output
			$x_k(t) = \sum \limits_{i=1}^{N} {c_{i}^k}^* \hat{\varphi}_i^k(t)$, $k=1,\cdots,K$. 
		\end{algorithmic}
	\end{algorithm}
	
	\subsection{\hl{Computational Complexity}}
	\hl{As discussed above, the core of 3D-SRMD lies in solving the BPDN problem. In each iteration of the SPGL1 algorithm used for this purpose, the primary cost is divided into two parts} \cite{van2009probing}. \hl{First, the matrix-vector multiplications involving the dense random feature matrix $\Psi$ and $\Psi^T$ incur a cost of $O(mKN)$. Second, projecting the current point onto the $\ell_1$-norm constraint ball, which employs a fast sorting heap structure, has a worst-case cost of $O(m\log m)$. Consequently, the total cost of 3D-SRMD is $O(mI(\log m+KN))$, where $I$ denotes the total number of SPGL1 iterations.}
	\section{Numerical Results}
	\hl{In this section, both simulated and real-world signals with crossover modes are considered to compare the performance of 3D-SRMD with other methods, i.e., NCMD and ICCD.}
	
	\subsection{Simulated Signal}
\begin{figure}[H]
		\centering
	
		\begin{minipage}[t]{0.5\linewidth}
			\centering
			\includegraphics[width=1.75in]{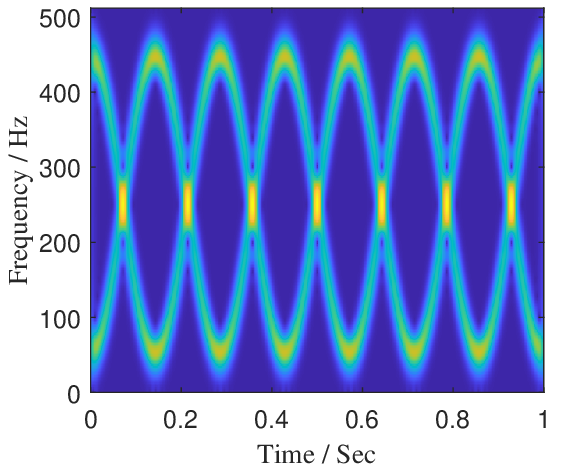}
			%\caption{fig2}
		\end{minipage}%
		\begin{minipage}[t]{0.5\linewidth}
			\centering
			\includegraphics[width=1.72in]{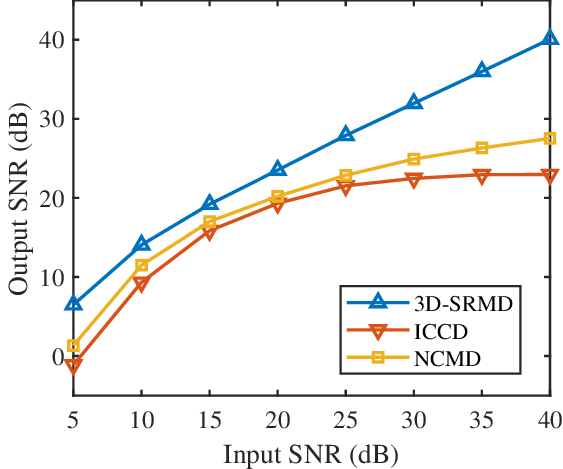}
		\end{minipage}
		\centering
		\caption{(\textbf{left}) The TF spectrogram of the simulated signal by STFT; (\textbf{right}) output SNR vs. input SNR for different methods}
		\label{fig1}
	\end{figure}

	Firstly, we consider the synthetic signal, as follows:
	\begin{equation}\label{sig1}
		s(t)=m_1(t)+m_2(t),
	\end{equation}
	with
	\begin{equation}
		\begin{aligned}
			m_1&(t)=\cos(2 \pi (250t-\frac{200}{7 \pi}\sin(7 \pi t))),
		\end{aligned}
	\end{equation}
	\begin{equation}
		\begin{aligned}
			m_2&(t)=\cos(2 \pi (250t+\frac{200}{7 \pi}\sin(7 \pi t))),
		\end{aligned}
	\end{equation}
	where the sampling frequency $f_s = 1024$ Hz and the time duration is $[0,1]$ s. The IFs of two modes exhibit oscillatory patterns and intersect at multiple time points (see Fig. \ref{fig1} (\textbf{left})).
	
	The parameters for each decomposition algorithm	involved in the comparison are set as follows. 
	Both NCMD and ICCD utilize the widely used IF estimation method RPRG \cite{chen2017separation}, which is capable of estimating crossover IFs in the 2D TF plane. To ensure fairness, NCMD employs the same noise estimation as 3D-SRMD.
	In 3D-SRMD, we set $\alpha=L/80$, $N=5000$, $\lambda=f_s/100$, and the maximum iteration count for the
	SPGL1 algorithm is set to 1000. In ICCD, a wider filter bandwidth should be chosen to improve the estimation accuracy of nonlinear oscillatory modes. Thus, we set $BW=f_s/10$, and the noise parameter is set to the default value of 5. Similarly, in NCMD, it is necessary to choose a larger penalty parameter to obtain a wider filter bandwidth, here set to $ 1 e - 2$.

	\begin{figure}[H]
		\centering
		\begin{minipage}[t]{0.5\linewidth}
			\centering
			\includegraphics[width=1.75in]{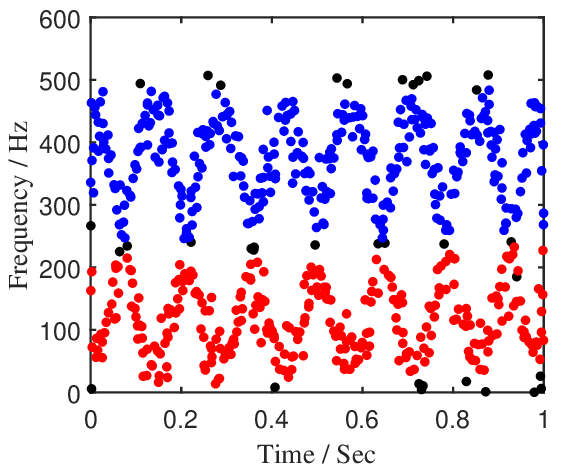}
			%\caption{fig1}
		\end{minipage}%
		\begin{minipage}[t]{0.5\linewidth}
			\centering
			\includegraphics[width=1.75in]{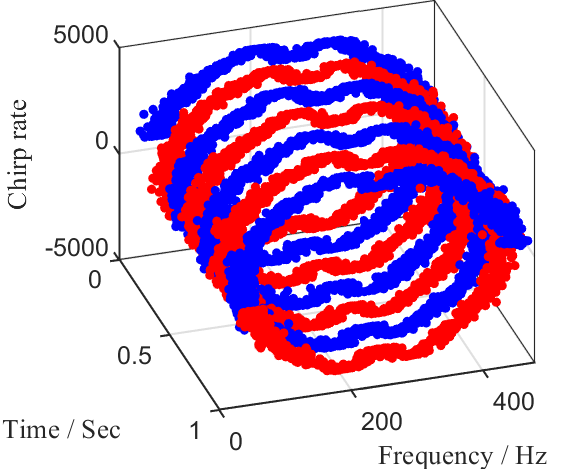}
			%\caption{fig2}
		\end{minipage}%
		\centering
		\caption{The random feature space of the simulated signal in (\textbf{left}) SRMD and (\textbf{right}) 3D-SRMD (different colors signify different modes).}
		\label{space}
	\end{figure}

	\begin{figure}[H]
		\centering
		\begin{minipage}[t]{0.5\linewidth}
			\centering
			\includegraphics[width=1.75in]{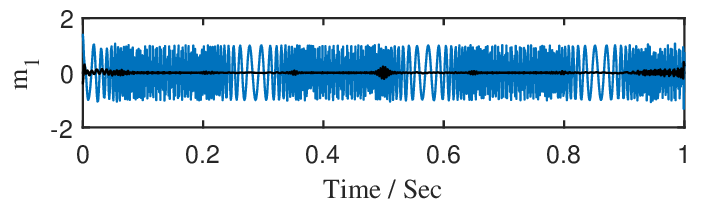}
			%\caption{fig2}
		\end{minipage}%
		\begin{minipage}[t]{0.5\linewidth}
			\centering
			\includegraphics[width=1.75in]{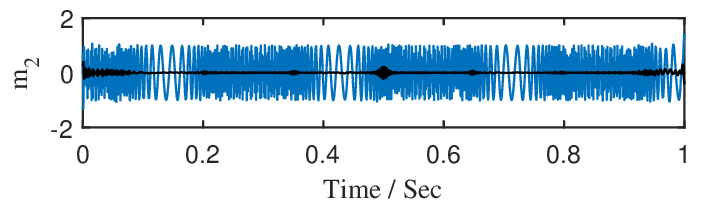}
			%\caption{fig2}
		\end{minipage}%
		
		\begin{minipage}[t]{0.5\linewidth}
			\centering
			\includegraphics[width=1.75in]{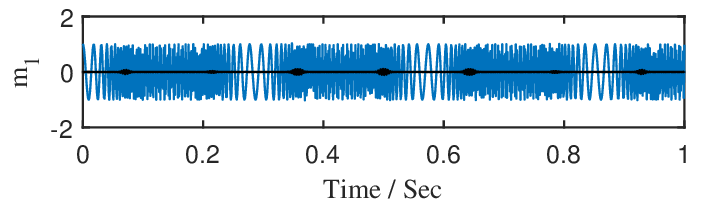}
			%\caption{fig2}
		\end{minipage}%
		\begin{minipage}[t]{0.5\linewidth}
			\centering
			\includegraphics[width=1.75in]{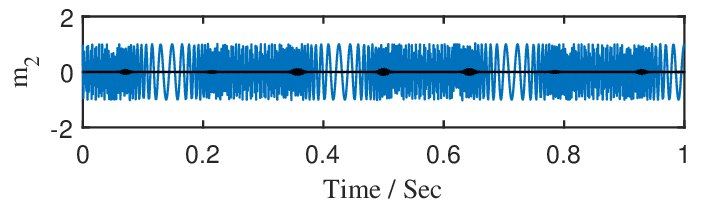}
			%\caption{fig2}
		\end{minipage}

		\begin{minipage}[t]{0.5\linewidth}
			\centering
			\includegraphics[width=1.75in]{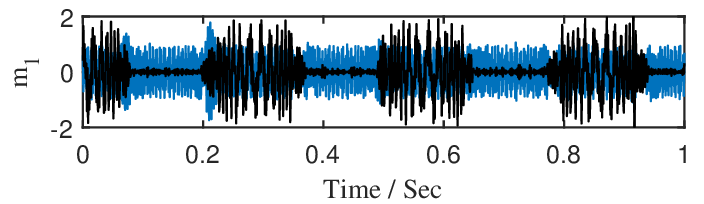}
			%\caption{fig2}
		\end{minipage}%
		\begin{minipage}[t]{0.5\linewidth}
			\centering
			\includegraphics[width=1.75in]{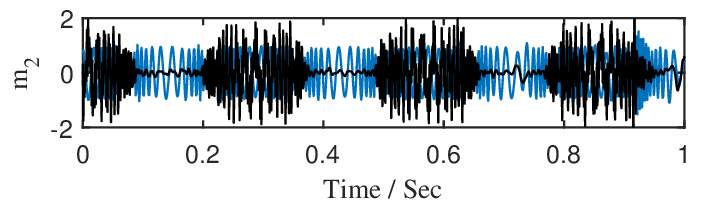}
			%\caption{fig2}
		\end{minipage}%
		
		\begin{minipage}[t]{0.5\linewidth}
			\centering
			\includegraphics[width=1.75in]{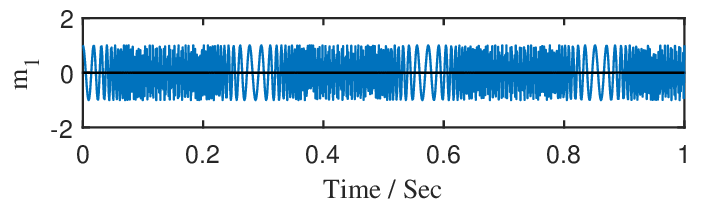}
			%\caption{fig2}
		\end{minipage}%
		\begin{minipage}[t]{0.5\linewidth}
			\centering
			\includegraphics[width=1.75in]{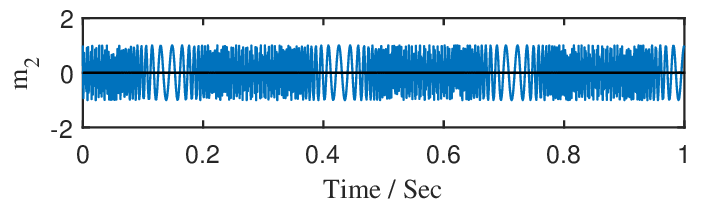}
			%\caption{fig2}
		\end{minipage}%
		
		\centering
		\caption{\hl{Analysis results (blue: estimated modes; black: estimation errors) for the simulated signal by ICCD (\textbf{First row}), NCMD (\textbf{Second row}), SRMD (\textbf{Third row}), (\textbf{Last row}) 3D-SRMD  (\textbf{Last row}). Note that the average SNR of the two signals by each method is 24.39 dB, 30.89 dB, 0.54 dB, and 55.70 dB, respectively.}}
		\label{fig2}
	\end{figure}
	
	Firstly, we evaluated the performance of each algorithm in a noise-free case. The decomposition results are shown in Fig. \ref{space} and Fig. \ref{fig2}.
	Due to the influence of oscillating IFs, ICCD experiences errors at crossover points and edges. NCMD exhibits some errors at the crossover point caused by the estimated errors in IFs. 
	SRMD reconstructs entirely erroneous modes, demonstrating its ineffectiveness in handling crossover IFs.
	Finally, the introduction of CR parameter enables 3D-SRMD to successfully  decompose crossover modes, presenting the best decomposition performance among these methods.
	
	To comprehensively evaluate the methods' performance, we also evaluated these methods under different noise levels. Note that the experiments are repeated 100 times at each noise level to obtain average results. As depicted in Fig. \ref{fig1} (\textbf{right}), 3D-SRMD exhibits superior performance at each noise level.
	
	\subsection{Real-World Signal}
\begin{figure}[H]
		\centering
		
		\begin{minipage}[t]{0.5\linewidth}
			\centering
			\includegraphics[width=1.75in]{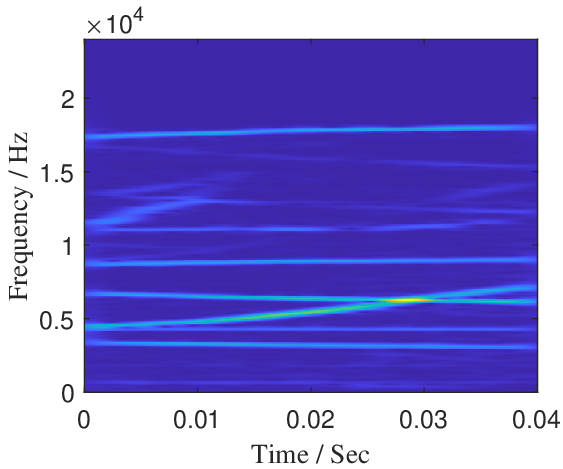}
			%\caption{fig2}
		\end{minipage}%
		\begin{minipage}[t]{0.5\linewidth}
			\centering
			\includegraphics[width=1.75in]{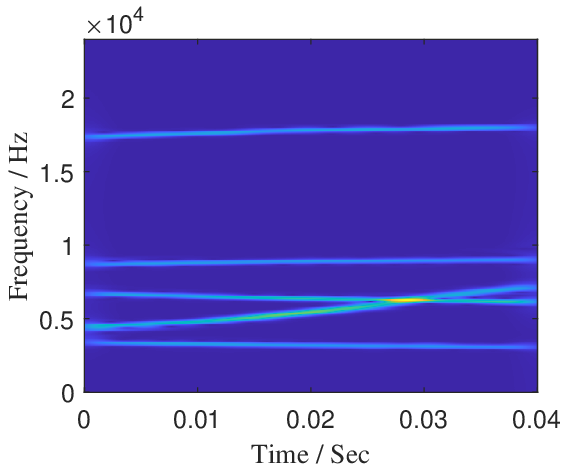}
			%\caption{fig2}
		\end{minipage}%	
		
		\begin{minipage}[t]{0.5\linewidth}
			\centering
			\includegraphics[width=1.75in]{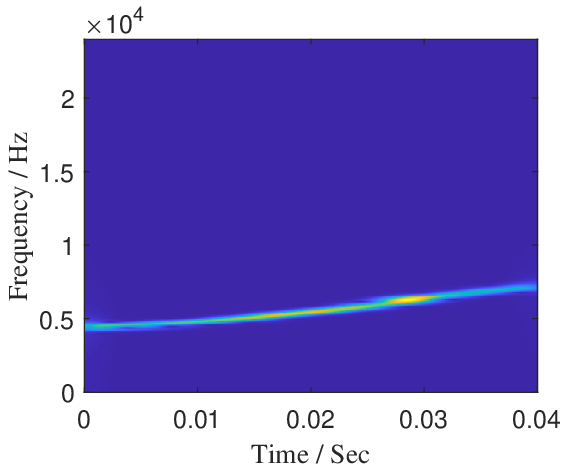}
			%\caption{fig2}
		\end{minipage}%
		\begin{minipage}[t]{0.5\linewidth}
			\centering
			\includegraphics[width=1.75in]{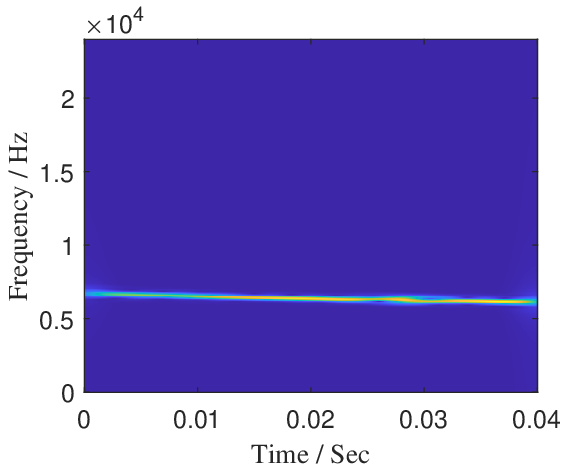}
			%\caption{fig2}
		\end{minipage}%
		
		\centering
		\caption{The whales' whistles and the decomposition results of the 3D-SRMD in STFT plane: (\textbf{top left}) the real signal; (\textbf{top right}) sum of the estimated modes; (\textbf{bottom}) the crossover frequency components.  }
		\label{fig4}
	\end{figure}

	To demonstrate the effectiveness of 3D-SRMD, a real-world signal from the whistles of the melon-headed whales \cite{voiceinthesea} is employed in this section.
	The sampling frequency is 48 kHz, while the time duration is 0.04 s. 
	The signal contains crossover frequency components and is contaminated by background noise (see in Fig. \ref{fig4} (\textbf{top left})).
	As exhibited in Fig. \ref{fig4}, 3D-SRMD successfully separate the modes with crossover IFs and significantly reduces the background noise, which indicates the potential of our method in practical applications.
	
	\section{Conclusion}
	
	This letter has proposed an advanced mode decomposition method called 3D-SRMD.
	By lifting the random feature space to the 3D TFC space, our method is capable of disentangling modes with crossover IFs.
	Furthermore, a concentrated distribution of random features is also developed to enhance separation accuracy and eliminate the need for clustering algorithms.
	Finally, 3D-SRMD demonstrates promising performance in both simulated and real-world signals. In the future, we will try to apply the 3D-SRMD in various areas, e.g., radar and biomedical systems. \hl{We will also aim to explore model compression techniques to optimize the computational efficiency of 3D-SRMD, enabling its adaptation to large-scale, real-time applications.}
	
	\bibliographystyle{IEEEtran}
	%\bibliography{IEEEabrv,}  
	\bibliography{Ref} 
\end{document}